\documentclass[prd,twocolumn,floatfix,amsmath,amssymb,floatfix]{revtex4}
\usepackage{graphicx,color,dcolumn,booktabs,bm}
\usepackage{longtable,lscape}
\usepackage{txfonts}
\usepackage{overpic}
\usepackage{amssymb}
\usepackage{indentfirst}
\usepackage{feynmf}   %{feynmp}
\usepackage{slashed}  %for Feynman symbols
\usepackage{cases}
\usepackage{color}
\usepackage{multirow}
\usepackage{epstopdf}
\usepackage{graphicx,color,dcolumn,booktabs,bm}

\graphicspath{{Figures/}} %
\usepackage[colorlinks,
                      citecolor=blue,
                      anchorcolor=red,
                      menucolor=red,
                      linkcolor=red,
                      filecolor=red,
                      runcolor=red,
                      urlcolor=blue,
                      frenchlinks=red]{hyperref}

\begin{document}

\title{Molecular components in the $J/\psi$ and the $\rho$-$\pi$ puzzle}
\author{Xing-Dao Guo$^{1,3}$}\email{guoxingdao@mail.nankai.edu.cn}
\author{Dian-Yong Chen$^{2,4}$}\email{chendy@seu.edu.cn}
\author{Xue-Qian Li$^3$}\email{lixq@nankai.edu.cn}
\author{Zhong-Yuan Yuan$^{1}$}
\email{yzy875@xzit.edu.cn}
\author{Shijing Sang$^{1}$}
\email{ssj876@xzit.edu.cn}

\affiliation{$^{1}$ College of Physics and New Energy, Xuzhou University of Technology, Xuzhou, 221111, P.R.~China}
\affiliation{$^{2}$ School of Physics, Southeast University, Nanjing, 211189, P.R.~China}
\affiliation{$^3$ Department of Physics, Nankai University, Tianjin, 300071, P.R. China}
\affiliation{$^4$ Lanzhou Center for Theoretical Physics, Lanzhou University, Lanzhou 730000, China}

\date{\today}

\begin{abstract}

Motivated by the large branching fractions of $J/\psi \to f_0 (1710) \omega/f_0(1710) \phi$ and the light exotic candidates, we find that there may exist molecular states composed of $f_0(1710) \omega$ and $f_0 (1710) \phi$, which correspond to $X(2440)$ and $X(2680)$ observed in a few decades before. The branching fraction of $X(2440)$ and $X(2680)$ to various $PV$ channels and $KK\omega(\phi)$ channels are estimated in the molecular scenario. In addition, the large branching fractions of $J/\psi \to f_0 (1710) \omega/f_0(1710) \phi$ indicate the sizable molecular components in the $J/\psi$ state. Thus, we consider the $J/\psi$ as the supperposition of $c\bar{c}(1S)$, $f_0(1710) \omega$ and $f_0 (1710) \phi$ molecular states, and these molecular components have significant impact on the light hadron decays of $J/\psi$, which may shield light on the long standing $\rho-\pi$ puzzle.

%Motivated by the experimental data $BR(J/\psi\to f_0(1710)\phi(\omega))>BR(J/\psi\to f_0(980)\phi(\omega))$ and recently observed light quark exotic states $X(2220)$, $X(2240)$ et al. in experiment. We suppose that there may exit molecular states $X_{f_0(1710)\phi}$ and $X_{f_0(1710)\omega}$ which may contribute to $J/\psi$ hadronic decays. Moreover, we suppose the experimental observed states $X(2440)$ and $X(2680)$ in a few decades before are constituted of $X_{f_0(1710)\phi}$ and $X_{f_0(1710)\omega}$ dominantly. In this work, we evaluate the branching ratios of the physical states $X(2440)$ and $X(2680)$ to various $PV$ and $KK\phi(\omega)$ final states to support our assumption. Since the corresponding experimental data lack or need to be updated, we lay our hope on the future more precise experiments such as BES III which will verify or negate our ansatz. We also suggest that the experimentally observed $J/\psi$ is not a pure $c\bar c$ state but a mixture of $c\bar c$ with two molecular states $X_{f_0\omega}$ and $X_{f_0\phi}$ which are responsible for the partial hadronic decay modes. Our numerical results indicate that if the effective coupling constant satisfies $g_X^2 g^2_{f_0}\sim 10^{0}$, the contribution of the molecular components can enhance the rate of most $J/\psi\to PV$ channels to the observed level. To determine the fractions of the molecular states in $J/\psi$, the criterion that the mass gaps among $\psi^\prime$, $\eta_c$ and $J/\psi$ remain in the experimental error tolerance ranges, is mandatory.

\end{abstract}

\maketitle

\section{Introduction}
Since the OZI suppressed light hadron pair decays of vector charmonia occur via three gluons annihilation, while their dilepton decays occur via a virtual photon. The estimation based the perturbative QCD indicated,
\begin{eqnarray}
	R={\mathcal{B}(\psi^\prime\to h)\over \mathcal{B}(J/\psi\to h)} =\frac{\mathcal{B}(\psi^\prime\to \ell^+ \ell^-)}{ \mathcal{B}(J/\psi\to \ell^+ \ell^-)} \sim 12\%,
\end{eqnarray}
which is referred to as the ``$12\%$ rule". Severe violation of ``$12\%$ rule" was firstly observed in the $\rho \pi$ channel, which was measured to be $(0.19 \pm 0.08)\%$ by Mark II Collaboration in the year of 1983~\cite{Franklin:1983ve}, and then observed in more channels. Such anomalous phenomena is named ``$\rho-\pi$ puzzle". In Table~\ref{tab1}, we collect the measured branching fractions of $J/\psi \to VP$ and $\psi^\prime \to VP$ and their ratios.

To solve the ``$\rho-\pi$ puzzle", various schemes have been proposed. In essence, there are two different major methods to address the discrepancy between the experimental measurements and the ``$12\%$ rule" expectation, it is introduce some addition mechanisms in the decays of either $\psi^\prime$ or $J/\psi $. For example, in Ref.~\cite{Rosner:2001nm}, the author suggested that $\psi^\prime$ is a $2S-1D$ mixing state rather than a pure 2S state, and the destructive interference greatly suppresses the branching fractions of $\psi^\prime\to \rho\pi$. Such kind of suppression could also be resulted from the possible final state interaction~ \cite{Li:1996yn,Brodsky:1997fj,Wang:2012mf,Zhao:2010zzv,Mo:2006cy,Zhao:2006gw,Suzuki:2000yq,Gu:1999ks,Tuan:1999ig,Chen:1998ma,Li:2007ky,Caldi:1975tx}. Whereas the estimations in some literatures suggested that the branching fractions of light hadron decays of $J/\psi$ is enhanced by some mechanisms. For example, Freund and Nambu~\cite{Freund:1975pn} considered that $J/\psi$ might mix with a $1^{--}$ glueball with the mass of $1.4\sim 1.8$ GeV which could also transit into $\rho \pi$, thus the rate of $J/\psi\to \rho\pi$ could be enhanced by the constructive interference~\cite{Brodsky:1987bb,Hou:1982kh,Hou:1982dy,Harris:1999wn,Anselmino:1993yg,Hou:1996qk,Chan:1999px,Chao:1996sf,Chaichian:1988kn}.

As one of important light hadron production platform, the $J/\psi$ decays have exhibited anomalous phenomena besides the ``$\rho-\pi$ puzzle". For example, the measured branching ratio of $J/\psi\to f_0(1710)\phi\to K\bar K\phi$ was measured to be $(3.6\pm0.6)\times 10^{-4}$ \cite{DM2:1988qci}, which is higher than the one for $J/\psi\to f_0(980)\phi$, which is  $(3.2\pm0.9)\times10^{-4}$~\cite{ParticleDataGroup:2022pth}. Similarly, one can notice that the branching ratio for $J/\psi\to f_0(1710)\omega\to K\bar K\omega$ is also greater than the one of $J/\psi\to f_0(980)\omega$, which are $(4.8\pm1.1)\times10^{-4}$~\cite{DM2:1988qci} and $(1.4\pm0.5)\times10^{-4}$ ~\cite{DM2:1988osw}, respectively.
Generally, the branching ratios for processes involving higher excited states are smaller than those only involving ground states in the $J/\psi$ decays due to the nodes effect and smaller phase space for the former processes. Thus, the larger $f_0(1710) \omega$ and $f_0 (1710) \phi$ branching ratios indicate anomalous strong coupling between $J/\psi$ and $f_0(1710) \omega/f_0(1710) \phi$. Additionally, the thresholds of $f_0(1710) \omega$ and $f_0 (1710) \phi$ are 2487 MeV and 2723 MeV, respectively.
In the vicinity of $f_0(1710) \omega$ threshold, a resonance $X(2440)$ with $M=2440\pm10$MeV, $\Gamma=310\pm20$ MeV~\cite{Rozanska:1979ub}have been reported, and recently, a resonance state around $2.4$GeV is observed in $\pi^+\pi^-\phi$ and $f_0(980)\phi$ invariant mass spectrums with quantum number $1^{--}$ \cite{BaBar:2007ptr,BES:2007sqy,Belle:2008kuo,Shen:2009mr,BESIII:2014ybv,BESIII:2021lho}.We suppose it may be the $X(2440)$ observed decades ago and assume it as a molecular state in this study.
In the vicinity of $f_0(1710) \phi$ threshold, a resonance $X(2680)$ with $M=2676\pm27$MeV, $\Gamma=150$ MeV~\cite{Caso:1970vv} have been reported.
Therefore, these two states could be molecular candidates of $f_0(1710) \omega$ and $f_0 (1710) \phi$, respectively.

Additionally, we also note that in Tab.\ref{tab1} the branching ratios of $\psi(2S)$ to various $PV$ final states are in the ranges of $(1.5\sim5.5)\times10^{-5}$ (expect $\psi(2S) \to \phi\pi$ ). While the branching ratios of $J/\psi$ to various $PV$ final states vary from $5.6\times10^{-3}$ to $8.1\times10^{-5}$(expect $J/\psi\to \phi\pi$ ).
Recently, $\mathcal{B}(\psi(3686)\to \phi K^0_SK^0_S)/\mathcal{B}(J/\psi\to \phi K^0_SK^0_S)=6.0\pm1.6\%$ is measured in experiment\cite{BESIII:2023pgb}. This ratio is also suppressed relative to the $12\%$ rule.
It seems like there is some kind of mechanistic effect the decays of $J/\psi$ to $PV$.
Thus we suppose that experimental observed state $J/\psi$ may contains tiny molecular state components which only effect some hadronic decay channels of $J/\psi$ but not effect others such as leptonic decays etc.
In this scenario, we expect a mixing scheme for the $J/\psi$ could shield light on the long standing ``$\rho-\pi$ puzzle".

\begin{table*}[htb]
\begin{center}
\caption{\label{tab1}The measured branching fractions of $J/\psi\to PV$ and $\psi^\prime\to PV$, where $P$ and $V$ refer to pseudoscalar and vector light meson, respectively.  ~\cite{ParticleDataGroup:2022pth}.}
\renewcommand\arraystretch{1.5}
\begin{tabular*}{160mm}{c@{\extracolsep{\fill}}ccccc}
\toprule[1pt]
  channel                    &  branch ratio                     &       channel                & branch ratio & ratio \\
\midrule[1pt]
$J/\psi \to \rho \pi$              & $(1.69\pm0.15)\times10^{-2}$       & $\psi^\prime \to \rho \pi$              & $(3.2\pm1.2)\times10^{-5}$  &    $(0.19\pm0.08)\%$  \\
$J/\psi \to \rho^0 \pi^0$          & $(5.6\pm0.7)\times10^{-3}$         & $\psi^\prime \to \rho^0 \pi^0$          &                             &    \\
$J/\psi \to K^{*+}\bar{K}^- +c.c.$ & $(6.0^{+0.8}_{-1.0})\times10^{-3}$ & $\psi^\prime \to K^{*+}\bar{K}^- +c.c.$ & $(2.9\pm0.4)\times10^{-5}$  & $(0.48\pm0.10)\%$  \\
$J/\psi \to K^{*0}\bar{K}^0 +c.c.$ & $(4.2\pm0.4)\times10^{-3}$         & $\psi^\prime \to K^{*0}\bar{K}^0 +c.c.$ & $(1.09\pm0.20)\times10^{-4}$     & $(2.60\pm0.54)\%$  \\
$J/\psi \to \omega \eta$           & $(1.74\pm0.20)\times10^{-3}$       & $\psi^\prime \to \omega \eta$           & $<1.1\times10^{-5}$         & $<(0.63\pm0.07)\%$  \\
$J/\psi \to \phi \eta $            & $(7.4\pm0.8)\times10^{-4}$         & $\psi^\prime \to \phi \eta$             & $(3.10\pm0.31)\times10^{-5}$     & $(4.19\pm0.62)\%$  \\
$J/\psi \to \phi \eta'$            & $(4.6\pm0.5)\times10^{-4}$         & $\psi^\prime \to \phi \eta'$            & $(1.54\pm0.20)\times10^{-5}$     & $(3.35\pm0.57)\%$  \\
$J/\psi \to \omega \pi$            & $(4.5\pm0.5)\times10^{-4}$         & $\psi^\prime \to \omega \pi^0$          & $(2.1\pm0.6)\times10^{-5}$  &  $(4.67\pm1.43)\%$ \\
$J/\psi \to \rho \eta$             & $(1.93\pm0.23)\times10^{-4}$       & $\psi^\prime \to \rho \eta $            & $(2.2\pm0.6)\times10^{-5}$  & $(11.40\pm3.40)\%$  \\
$J/\psi \to \phi \pi$              & $3\times10^{-6}$                   & $\psi^\prime \to \phi \pi^0$            & $<4\times10^{-7}$           & $<13.33\%$  \\
$J/\psi \to \omega \eta'$          & $(1.89\pm0.18)\times10^{-4}$       & $\psi^\prime \to \omega \eta'$          & $3.2^{+2.5}_{-2.1}\times10^{-5}$ & $(16.93\pm13.33)\%$  \\
$J/\psi \to \rho \eta'$            & $(8.1\pm0.8)\times10^{-5}$         & $\psi^\prime \to \rho \eta'$            & $1.9^{+1.7}_{-1.2}\times10^{-5}$ & $(23.46\pm21.12)\%$  \\
\bottomrule[1pt]

%\bottomrule
\end{tabular*}
\vspace{0mm}
\end{center}
\end{table*}

In this work, we propose that the experimentally observed $J/\psi$ is a mixture of $c\bar c$ and the hadronic molecules $X_{f_0(1710)\omega}$ and $X_{f_0(1710)\phi}$ (hereinafter we denote as $X_{f_0\omega}$ and $X_{f_0\phi}$ respectively for short). In the $J/\psi$ light hadron decays, the molecular components break down into on-shell $f_0(1710)$ and $\phi(\omega)$, which then transition into light hadron pairs by exchanging an appropriate light hadron. It is important to note that the exchanged light hadron is also on-shell, potentially increasing the rescattering contributions. Consequently, although the proportion of hadronic molecules $X_{f_0\phi}$ and $X_{f_0\omega}$ in $J/\psi$ state may be small, they still have a significant impact on the light hadron decays.

This work is organized as follows. After the introduction, we analyze the mixing effect between $c\bar c$ and hadronic molecule in section II. Then we roughly estimate the fractions of $X_{f_0\phi}$ and $X_{f_0\omega}$ components in $J/\psi$ in section III. In section IV, we calculate branching fractions of $J/\psi\to VP$ as long as the components $X_{f_0\phi}$ and $X_{f_0\omega}$ are taken into account. The last section is devoted to a short summary.

\section{Spectra of $c\bar{c}$ bound state, the molecules and their mixing}

Based on the ansatz that the physical states $X(2440)$, $X(2680)$ and $J/\psi$ are the mixtures of the molecular states $X_{f_0\omega}$, $X_{f_0\phi}$ and the charmonium $c\bar c(1S)$ through a unitary matrix $U$ transformation, we have the relation as follows,
\begin{eqnarray}
\left(
  \begin{array}{ccc}
    |X(2440)\rangle \\
    |X(2680)\rangle \\
    |J/\psi\rangle \\
\end{array}
\right) = U \left(
  \begin{array}{ccc}
    |X_{f_0\omega }\rangle \\
    |X_{f_0\phi }\rangle   \\
    |c\bar c(1S)\rangle    \\
  \end{array}
\right)
\label{pstoqs}
\end{eqnarray}
where $U$ is a unitary matrix with the compact form as,
\begin{eqnarray}
 U=\left(
  \begin{array}{ccc}
    c_{11} & c_{12} & c_{13} \\
    c_{21} & c_{22} & c_{23} \\
    c_{31} & c_{32} & c_{33}\\
    \end{array}
\right).
\end{eqnarray}
Then one has,
\begin{eqnarray}
|J/\psi\rangle= c_{31} | X_{f_0 \omega}\rangle + c_{32} | X_{f_0 \phi}\rangle  + c_{33} | c\bar{c}(1S)\rangle. \label{Eq:Mixing}
\end{eqnarray}
For $\psi^\prime$, its mass is far away from the ones of the molecular states $X_{f_0 \omega}$ and $X_{f_0 \phi}$, thus, the mixing between $c\bar{c} (2S)$ and the molecular states should be ignorable. One can suppose that the magnitudes of $c\bar{c}(1S) \to PV $ and $c\bar{c}(2S) \to PV $ satisfy the ``$12\%$ rule", while the violations is at least partly resulted from the mixing of the charmonium and light meson-meson molecular states. In order to estimate the mixing between the charmonium and the molecular states, we first investigate the spectra of charmonium and the molecular states individually.

\subsection{Spectrum of the $c\bar c$ bound state}

Following Ref.~\cite{Godfrey:1985xj} one obtains the mass spectra of charmonia by solving Schr\"{o}dinger equation with one-gluon-exchange plus a linear confinement potential which manifests the non-perturbative QCD effects. The   Hamiltonian for $1^{--}$ states is,
\begin{eqnarray}
H=H_0 +H'
\end{eqnarray}
with
\begin{eqnarray}
H_0=\sum\limits^2_{i=1} \left(m_i+\dfrac{p^2}{2m_i}\right)+\dfrac{-4}{3}\dfrac{\alpha_s (r)}{r} +\kappa r + c
\end{eqnarray}
and
\begin{eqnarray}
H'=\dfrac{1}{4}\dfrac{32\pi}{9m_1 m_2}\alpha_s(r)\left(\dfrac{\sigma}{\sqrt\pi}\right)^3 e^{-\sigma^2 r^2}
\end{eqnarray}
where $\kappa=0.18$ GeV$^2$, $\sigma=3.0996$ GeV and $m_c=1.628$ GeV~\cite{Godfrey:1985xj}. The constant $c$ is zero-point energy which is fixed by fitting the experimental data. From above Hamiltonian, one obtains the mass of $1S$ and $2S$ state, while their mass difference is independent of the constant $c$. By setting $\psi(2S)$ to be a pure $c \bar{c}(2S)$ state, i.e.,  $m_{c \bar{c}(2S)}=(m_{\psi(2S)})_{\mathrm{exp}}$, one has, $c=(448.93 \pm 0.06)$ MeV. With this value of $c$, one has, $m_{c\bar c(1S)}=(3088.19 \pm 0.06)$ MeV, which is about 10 MeV below the PDG average~\cite{ParticleDataGroup:2022pth}. Additionally, the masses of  ground and the first excited state of $c\bar c(0^-)$ states are $m_{\eta_c}=3022.02$ MeV and $m_{\eta_c(2S)}=3641.90$ MeV, respectively.

\begin{figure}[t]
\includegraphics[width=6cm]{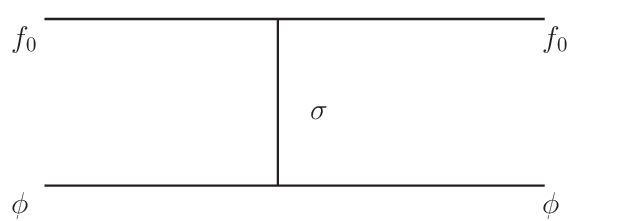}
\caption{
The Feynman diagram for $f_0(1710)\phi$ scattering by exchanging a $\sigma$ meson. \label{allowed}}
\end{figure}

\subsection{Masses of light meson-meson bound states $X_{f_0\omega}$ and $X_{f_0\phi}$}

The mass spectra the light meson-meson bound states can also be estimated by solving the non-relativistic Schr\"{o}dinger equation where the potentials between the meson components could be induced by one-boson exchange as show in Fig.~\ref{allowed}~\cite{Chen:2019uvv,Zhao:2013ffn}.
In Refs.\cite{Cheng:2002ai,Cheng:2005nb} the authors consider $\sigma$ and $f_0(980)$ as mixture of $(u\bar u+d\bar d)/\sqrt2$ and $s\bar s$, thus we consider that $\sigma$ can couple with $f_0(1710)$ and $\phi$.
Taking $f_0(1710) \phi$ system as an example, one can transform the scattering amplitude of the elastic scattering process in the momentum space into a non-relativistic effective potential in coordinate space. To obtain the amplitude corresponding to diagram in Fig.~\ref{allowed}, the following effective Lagrangians are employed,
\begin{eqnarray}
&\mathcal{L}_{f_0 f_0 \sigma}= g_\sigma m_{f_0} \langle f_0 f_0 \sigma \rangle\\
&\mathcal{L}_{\phi \phi \sigma}= g_\sigma m_{\phi} \langle \phi \phi \sigma \rangle.
\end{eqnarray}
Unlike the hadronic scattering where incoming and out going mesons are on their mass shells, the ingredient mesons in the bound state are off-shell, thus, a form factor which partly compensates the off-shell effect is introduced, which is in the form,
\begin{eqnarray}
\mathcal{F} (q^2, m^2_\sigma)=\frac{\Lambda^2-m^2_\sigma}{\Lambda^2-q^2}
\end{eqnarray}
where $m_\sigma$ and $q$ are the mass and four momentum of exchanged $\sigma$ meson, respectively. Then the resultant effective potential reads as~\cite{Chen:2019uvv}
\begin{eqnarray}
V(\Lambda, m_\sigma, r) =-g^2_\sigma \left[\dfrac{1}{4\pi r}\left(e^{-m_\sigma r} -e^{-\Lambda r}\right) - \dfrac{\Lambda^2 -m_\sigma^2}{8\pi \Lambda} e^{-\Lambda r} \right].
\end{eqnarray}
Following Ref.~\cite{Chen:2019uvv} the relation $g_\sigma=2/3 g_{\sigma NN}$ and $g_{\sigma NN}^2/4\pi=5.69$ are assumed.
However, we should note that in Ref.~\cite{Chen:2019uvv} the mesons in corresponding effective Lagrangian are all ground states, namely $f_0(980)$.
But $f_0(1710)$ is a higher excited state in $f_0$ family.
For the $\sigma$ exchanging, the interaction between heavy quarks and light quarks is ignored in general, namely people only consider the light quarks contributions\cite{Wang:2019aoc}.
Thus for meson scattering the effective coupling constant is $2/3$ times smaller than that from baryons, this is suitable for $f_0(1710)$ also. Moveover, Following ref.~\cite{Cao:2010km,Ouyang:2009kv} we have $g^2/4\pi=3.20$, $4.45$ and $0.085$ for three $N^*(1440)N\sigma$, $N^*(1680)N\sigma$ and $N^*(1710)N\sigma$ vertices respectively.
Thus to give a rough estimate, we adopt $g_\sigma(f_0(1710))\sim g_\sigma(f_0(980))$ here.
Following Ref.~\cite{Zhao:2013ffn} we take $\Lambda=1.5\sim2.0$GeV.

After taking $m_{f_0(1710)}=1704$MeV and $m_\phi=1019$MeV, we get $m_{X_{f_0\phi}}=2647\sim 2701$MeV.
Adopting similar method, we obtain $m_{X_{f_0\omega}}=2440\sim2477$MeV.

\subsection{Mixing of charmonium and molecular states}

In the present estimation,  we impose the following conditions for the unitray matrix as the mandatory conditions, the determinant of the matrix should be unity and all the matrix elements are real. As indicated in Eq. (\ref{pstoqs}), the unitary matrix $U$  transforms the unphysical states $|X_{f_0\omega}\rangle, |X_{f_0\phi}\rangle$ and $|c\bar c(1S)\rangle$ into the physical eigenstates $|X(2440)\rangle, |X(2680)\rangle$ and $|J/\psi\rangle$, and at the same time diagonalizes the mass matrix $\tilde M_q$ as
\begin{eqnarray}
M_{\mathrm{mass}}=U \tilde M_q U^\dagger
\end{eqnarray}
with
\begin{eqnarray}
M_{\mathrm mass}=
\left(
  \begin{array}{ccc}
    m_{X(2440)} & 0 & 0 \\
    0 & m_{X(2680)} & 0 \\
    0 & 0 & m_{J/\psi} \\
    \end{array}
\right),
\end{eqnarray}
and
\begin{eqnarray}
\tilde M_q=
\left(
  \begin{array}{ccc}
    m_{X_{f_0\omega}} & \lambda_1 & \lambda_2 \\
    \lambda_1 & m_{X_{f_0\phi}} & \lambda_3 \\
    \lambda_2 & \lambda_3 & m_{c\bar c(1S)} \\
    \end{array}
\right).
\end{eqnarray}
Namely, $m_{X(2440)},\: m_{X(2680)}$, and $m_{J/\psi}$ are the three roots of equation,
\begin{widetext}
\begin{eqnarray}
&&m^3 - m^2\left(m_{c\bar c(1S)}+m_{X_{f_0\phi}}+m_{X_{f_0\omega}}\right)
+m\left(m_{c\bar c(1S)} m_{X_{f_0\omega}}+ m_{c\bar c(1S)} m_{X_{f_0\phi}}+ m_{X_{f_0\omega}} m_{X_{f_0\phi}}
-\lambda_1^2- \lambda_2^2- \lambda_3^2\right)+\nonumber\\
&&\qquad \qquad \qquad \left(\lambda_1^2 m_{c\bar c(1S)}+ \lambda_2^2 m_{X_{f_0\phi}}+ \lambda_3^2 m_{X_{f_0\omega}}-
2\lambda_1\lambda_2\lambda_3- m_{X_{f_0\omega}} m_{X_{f_0\phi}} m_{c\bar c(1S)}\right)
=0.
\end{eqnarray}
\end{widetext}
Generally, we have three unknown variables in the hermitian matrix $\tilde M_q$, which is $\lambda_1,\; \lambda_2$ and $\lambda_3$. While there are three independent equations by which we fix all these three unknown variables.
In principle, we could simultaneously fix the values of three non-diagonal matrix element by setting the physical masses of $m_{X(2440)},\: m_{X(2680)}$, and $m_{J/\psi}$ as the eigen-values of the mass matrix. However, we notice that the secular equation cannot be solved in normal way, so that we adopt an alternative method to obtain the ranges of three non-diagonal matrix elements. We pre-determine the ranges of the elements of the unitary transformation matrix $U$ which diagonalizes the mass matrix $\tilde M_q$ and then substitute them into the secular equation to check if the equation can be satisfied, namely if all the requirements (unitarity, etc.) are fulfilled. Then, we can obtain the unitary transformation matrix $U$ and the mass matrix $\tilde M_q$ as,
\begin{widetext}
\begin{eqnarray}
 U=\left(
  \begin{array}{ccc}
    -0.990\sim-0.985 & -0.147\sim-0.049 & -0.099\sim0.141 \\
    0.056\sim0.159  & -0.976\sim-0.997 & 0.050\sim0.150 \\
    0.061\sim0.133  & 0.057\sim0.161   & 0.985\sim0.990 \\
\end{array}
\right)
\label{18}
\end{eqnarray}
and
\begin{eqnarray}
 M_q=\left(
  \begin{array}{ccc}
    2438\sim2463 & -33\sim-7 & 45\sim86 \\
    -33\sim-7 & 2645\sim2708 & 24\sim74 \\
    45\sim86 & 24\sim51 & 3079\sim3087 \\
\end{array}
\right) ~\mathrm{MeV}
\end{eqnarray}
\end{widetext}

From the above two matrices, we find that the pre-determined ranges of the mass of $X_{f_0\omega}$ and $X_{f_0\phi}$ are $(2438\sim 2463)$ MeV and $(2645 \sim 2708)$ MeV. respectively, which are very close to the ones obtained by the potential model estimations. Mover, the value $m_{c\bar{c}(1S)}$ is fitted to be $(3079\sim3087)~\mathrm{MeV}$, which is also very close to the one obtained from quark model, which is $(3088.19 \pm 0.06)$ MeV.

\begin{figure}[htb]
\begin{center}
\includegraphics[width=8.5cm]{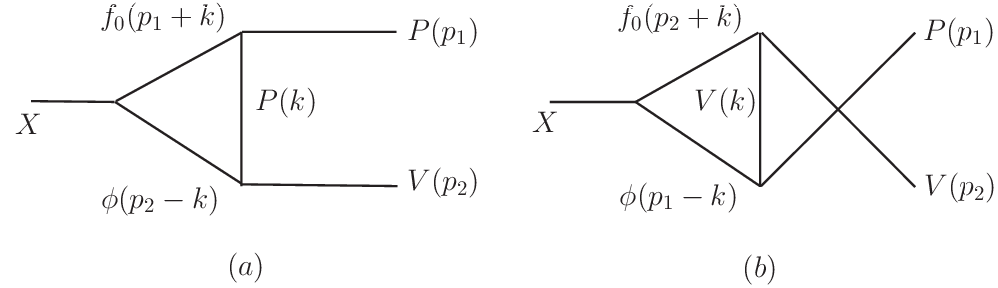}
\caption{
\label{fig2}
The Feynman diagrams contributing to $X_{f_0\phi} \to P V$. }
\end{center}
\end{figure}

\section{Molecular components contributions to $PV$ channels}

Besides the contributions from the $c\bar{c}$ annihilation, the molecular components in the $J/\psi$ should also have important effect to the decay $J/\psi \to PV$. For the molecular state, its components transit into light pseoduscalar and vector mesons exchanging a proper light meson, such kind of meson loop connect the molecular states and the final states.  Taking $|f_0 \phi\rangle$ as an example, the Feynman diagrams contributed to the decays $X_{f_0 \phi} \to PV$ are presented in Fig.~\ref{fig2}. In the present work, these diagram are estimated in the hadron level, and the relevant vertices read~\cite{Chen:2019uvv}
\begin{eqnarray}
\langle X_{f_0 V} | f_0 V\rangle &=& g_X \;\Lambda_0\; X_{f_0 V} V f_0 ,\nonumber\\
\langle f_0 | PP\rangle &=& g_{f_0} m_P\; \mathrm{Tr}[ PP] f_0,\nonumber\\
\langle f_0 | VV\rangle &=& g_{f_0} m_V\;  \mathrm{Tr}[ VV] f_0,\nonumber\\
\langle\phi | PV\rangle &=& \frac{G}{\sqrt2}\;\varepsilon^{\lambda\nu\alpha\beta} \mathrm{Tr}[ \partial_\lambda V_\nu \partial_\alpha V_\beta P ],
%g_{\phi K K^*}\varepsilon^{\lambda\nu\alpha\beta}p_\phi^\lambda \varepsilon_\phi^\nu (p_\phi-k_{K^*})^\alpha\varepsilon_{K^*}^\beta,
\label{19}
\end{eqnarray}
$\Lambda_0=1$GeV is dimensional parameter to make sure the effective coupling constant $g_X$ dimensionless, $m_V$ and $m_P$ denote the masses of vector and pseudoscalar mesons, respectively. The matrix form of the psedoscalar and vector mesons are
\begin{eqnarray}
&P
=
\left(
  \begin{array}{ccc}
\frac{\pi^0}{\sqrt2}+ \frac{\eta_8}{\sqrt6}+ \frac{\eta_1}{\sqrt3} & \pi^+ & K^+ \\
\pi^- & -\frac{\pi^0}{\sqrt2}+ \frac{\eta_8}{\sqrt6}+ \frac{\eta_1}{\sqrt3} & K^0 \\
K^- & \bar K^0 & -\frac{2\eta_8}{\sqrt6}+ \frac{\eta_1}{\sqrt3} \\
  \end{array}
\right)\nonumber
\\
\nonumber\\
&V
=
\left(
  \begin{array}{ccc}
\frac{\rho^0}{\sqrt2}+ \frac{\omega_8}{\sqrt6}+ \frac{\omega_1}{\sqrt3} & \rho^+ & K^{*+} \\
\rho^- & -\frac{\rho^0}{\sqrt2}+ \frac{\omega_8}{\sqrt6}+ \frac{\omega_1}{\sqrt3} & K^{*0} \\
K^{*-} & \bar K^{*0} & -\frac{2\omega_8}{\sqrt6}+ \frac{\omega_1}{\sqrt3} \\
 \end{array}
\right)
.\nonumber
\label{}
\end{eqnarray}
with mixing parameters to be,
\begin{eqnarray}
\eta_8&=& \eta\cos\theta+ \eta'\sin\theta,\nonumber\\
\eta_1&=& -\eta\sin\theta+ \eta'\cos\theta,\nonumber\\
\omega_8&=& \omega\cos\varphi+ \phi\sin\varphi,\nonumber\\
\omega_1&=& -\omega\sin\varphi+ \phi\cos\varphi,
\end{eqnarray}
with the mixing angle $\theta$  and $\varphi$ to be $\sin\theta=-0.31\pm0.11$~\cite{ParticleDataGroup:2022pth} and $\sin\varphi=-0.76$~\cite{Chen:2011cj}.

With the above vertices, we can obtain the amplitudes correspond to the diagram (a) and (b) in Fig. ~\ref{fig2}, which are,
%\begin{widetext}
\begin{eqnarray}
\mathcal{M}_a&=&\displaystyle\int\frac{d^4k}{(2\pi)^4}
\frac{g_{f_0} m_P}{(p_1+k)^2- m^2_{f_0} }
\frac{g_X g^{\mu\nu} \varepsilon_X^\mu }{(p_2-k)^2- m^2_\phi }
\nonumber\\
&&\times \dfrac{g_{\phi PV} \varepsilon^{\lambda\nu\alpha\beta} (p_2-k)^\lambda p_2^\alpha \varepsilon^{*^\beta}_V }{k^2-m_P^2}\nonumber\\
\mathcal{M}_b&=&\displaystyle\int\frac{d^4k}{(2\pi)^4}
\dfrac{g_{\phi PV} \varepsilon^{\lambda\nu\alpha\sigma} (p_1-k)^\lambda k^\alpha }{(p_1-k)^2- m^2_\phi }
\dfrac{g_X g^{\mu\nu} \varepsilon_X^\mu }{(p_2+k)^2- m^2_{f_0} }\nonumber\\
&\times &
\dfrac{g_{f_0} m_V g^{\sigma\beta} \varepsilon^{*^\beta}_V }{k^2-m_V^2},
\label{Eq:Amp}
\end{eqnarray}
respectively.

\section{Numerical Results and discussions}

\subsection{Coupling Constants}
Consider $SU(3)$ symmetry, the coupling constants in the hadron vertex in Eq. (\ref{19}) satisfy $G=(3g^2)/(4\pi^2 f_\pi)$, $f_\pi=93$~MeV and $g=(12)/(2\sqrt2)$\cite{Chen:2019uvv}. In the mixing scenario, $X(2240)$ and $X(2680)$ are molecular state with a very small $c\bar{c}(1S)$ component, and the $c\bar{c}$ components decays into light $PV$ is further suppressed by OZI suppression rule, thus when we estimate the $X(2240)/X(2680) \to PV$, the $c\bar{c}(1S)$ components could be ignored and $X(2240)/X(2680)$ could be considered as pure $f_0(1710) \omega/f_0(1710) \phi$ molecular state. In the molecular scenario, the coupling constant of the molecular state and its components could be estimated by \cite{Weinberg:1962hj,Cleven:2014oka},
\begin{eqnarray}
g_{X}^2 = \dfrac{16\pi(m_{f_0(1710)}+ m_V)^2}{\mu}\sqrt{2\mu E_B} \dfrac{1}{ \Lambda_0^2}. \label{Eq:gX}
\end{eqnarray}
with $\mu=(m_{f_0(1710)} m_V)/(m_{f_0(1710)}+ m_V)$ and $E_B=m_{f_0(1710)}+ m_V-m_{X}$. Taking $m_{f_0(1710)}=1704$ MeV, $m_\omega=783$ MeV, $m_\phi=1019$ MeV, $m_{X_{f_0\omega}}=(2440\sim 2477)$ MeV and $m_{X_{f_0\phi}}=(2647 \sim 2701)$ MeV, we get $g_{X_{f_0\omega}}=9.6\pm1.8$ and $g_{X_{f_0\phi}}=11.7\pm1.8$.

As for the value of the coupling constant $g_{f_0}$, by reproducing the experimental data $\Gamma(f_0(1710))=123$ MeV~\cite{ParticleDataGroup:2022pth} and $\mathcal{B}(f_0(1710) \to KK)= 0.38\pm0.19$~\cite{Longacre:1986fh}, $\mathcal{B}(f_0(1710) \to \eta\eta)= 0.22\pm0.12$~\cite{Albaladejo:2008qa} and $\mathcal{B}(f_0(1710) \to \pi\pi)= 0.039\pm0.024$~\cite{Longacre:1986fh}, we have $g_{f_0}=2.24\pm 0.56$, $g_{f_0}=3.17\pm 0.86$ and $g_{f_0}=4.68\pm 1.44$ respectively. As a matter of fact~\cite{pdglive}, all the three experimental values are not reliably measured, and the value of $g_{f_0}$ obtained from different experimental data are not very consistent with each other. Thus, in our calculation we just consider $g_{f_0}$ as a free parameter for the first step.

It should be noted that some channels in Table~\ref{tab1} are isospin violated, thus, in the present work, some isospin violated vertices are also included. From the branching fraction $\mathcal{B}(\phi \to \omega\pi)_{exp}= (4.7\pm0.5)\times10^{-5}$, we obtain, $g_{\phi\omega\pi}=0.04$. As for the other involved isospin violation vertices, for example $g_{\phi\rho\eta}$, $g_{\phi\rho\eta'}$, $g_{\omega\rho\eta}$, $g_{\omega\rho\eta'}$ and $g_{\omega\omega\pi}$, since the corresponding experimental data are lack, here we consider  that they are in the same orders as $g_{\phi\omega\pi}$ and take $g_{\phi\rho\eta}=g_{\phi\rho\eta'}=g_{\omega\rho\eta}=g_{\omega\rho\eta'}=g_{\omega\omega\pi}=0.04$ in the present calculations.

\subsection{The branching fractions of $X(2240)/X(2680)\to PV$}

As we discussed in the above subsection, the coupling constant $g_{f_0}$ is considered as a free parameter for the first step.
The partial width of $X(2440) \to VP$ satisfy
\begin{eqnarray}
\Gamma(X(2440) \to VP)_{\mathrm{Theory} } &=& \frac{1}{2\;m_{X(2440)} } \int d\Pi_2\nonumber \\
&& |c_{11} \; \mathcal{M}(X_{f_0\omega} \to VP)\nonumber\\
&& + c_{12}\; \mathcal{M}(X_{f_0\phi} \to VP)\\
&& + c_{13}\; \mathcal{M}(c\bar c(1S) \to VP)|^2\nonumber,
\label{Eq:24402PV}
\end{eqnarray}
with the amplitude for $c \bar c(1S)\to PV$ is
\begin{eqnarray}
\mathcal{M}(c\bar c(1S) \to VP)=g_{c\bar c PV} \varepsilon^{\lambda\nu\alpha\beta}
 p_\psi^\lambda \varepsilon^\nu_\psi p_V^\alpha \varepsilon^{*^\beta}_V .
\end{eqnarray}
The constants $g_{c\bar c PV}$ are obtained by fitting $\mathcal{B}(c\bar c(1S)\to PV)_{\mathrm{Exp}}$ data with the relation
\begin{eqnarray}
\mathcal{B}(c\bar c(1S)\to PV)_{\mathrm{Exp}}= {\mathcal{B}(\psi^\prime\to PV)_{\mathrm{Exp}}}/{(13.3\%\times c_{33}^2)},
\end{eqnarray}
where $13.3\%$ comes from $\mathcal{B}(\psi'\to l^+l^-)_{\mathrm{ Exp}}/\mathcal{B}(J/\psi\to l^+l^-)_{\mathrm{ Exp}}$ with $l=e,\;\mu$. In Table \ref{tab3} we list corresponding data and values.
\begin{table*}[htb]
\begin{center}
\caption{\label{tab3}
 The experimental data of $J/\psi$ and $\psi'$ to lepton pair\cite{ParticleDataGroup:2022pth}.}
\renewcommand\arraystretch{1.5}
\begin{tabular}{p{3cm}<\centering p{3cm}<\centering p{2cm}<\centering p{3cm}<\centering p{2cm}<\centering }
\toprule[1pt]
  decay channel         &  branch ratio                   &       decay channel    &  branch ratio                & ratio           \\
\midrule[0.5pt]
$J/\psi \to e^+e^-$     & $ (5.971\pm0.032)\times10^{-2}$ & $\psi' \to e^+e^-$     & $(7.93\pm0.17)\times10^{-3}$ &  $(13.28\pm0.29)\%$  \\
$J/\psi \to \mu^+\mu^-$ & $(5.961\pm0.033)\times10^{-2}$  & $\psi' \to \mu^+\mu^-$ & $(8.0\pm0.6)\times10^{-3}$   & $(13.42\pm1.01)\%$  \\
\toprule[1pt]
\end{tabular}
\vspace{0mm}
\end{center}
\end{table*}
The partial width of $X(2680) \to VP$ similar and we omit it for simplification.

Taking $\Gamma_{X(2440)}=310$ MeV and $\Gamma_{X(2680)}=150$ MeV~\cite{ParticleDataGroup:2022pth}, one  can obtain the branching fractions of $X(2240)/X(2680) \to PV$.
Now let us briefly discuss the effective coupling constant $g_{f_0}$.
Although its value is not well measured, the roughly range of it is $2\sim4$, here we take $g_{f_0}=2$ and $3$ as examples and list the estimated results in Table \ref{tab4}.

\begin{table*}[htb]
\begin{center}
\caption{\label{tab4}
The branching fractions of $X(2440)$ and $X(2680)$ decaying into a light pseudoscalar and vector meson, where we take $g_{f_0}=2$ and $3$ as examples.
The branching fractions of the decay channels $X(2440)\to \omega K\bar K$ and $X(2680)\to \phi K\bar K$ are also listed. The uncertainties of the present estimations are resulted from $c_{ij}$.}
\renewcommand\arraystretch{1.5}
\begin{tabular}{p{2cm}<\centering p{3cm}<\centering p{3cm}<\centering p{3cm}<\centering p{3cm}<\centering}
\toprule[1pt]
      Channel     &  \multicolumn{2}{c}{$\mathcal{B}(X(2440)\to PV)_{theory}$} &  \multicolumn{2}{c}{$\mathcal{B}(X(2680)\to PV)_{theory}$}\\
\midrule[0.5pt]
                 ~&~ $g_{f_0}=2$              ~&~ $g_{f_0}=3$             ~&~ $g_{f_0}=2$              ~&~ $g_{f_0}=3$  \\
\midrule[1pt]
$\rho^0 \pi^0$    & $(7.97\pm0.14)\times10^{-3}$ & $(1.78\pm0.03)\times10^{-2}$& $(1.45\pm0.73)\times10^{-3}$ & $(3.21\pm1.59)\times10^{-3}$ \\
$K^{*+}\bar{K}^-$ & $(1.05\pm0.18)\times10^{-3}$ & $(2.31\pm0.40)\times10^{-3}$& $(1.29\pm0.09)\times10^{-2}$ & $(2.88\pm0.19)\times10^{-2}$\\
$K^{*0}\bar{K}^0$ & $(1.13\pm0.20)\times10^{-3}$ & $(2.42\pm0.43)\times10^{-3}$& $(1.31\pm0.10)\times10^{-2}$ & $(2.92\pm0.21)\times10^{-2}$\\
$\omega \eta$     & $(2.89\pm0.13)\times10^{-4}$ & $(6.22\pm0.25)\times10^{-4}$& $(1.35\pm0.51)\times10^{-4}$ & $(2.85\pm1.04)\times10^{-4}$\\
$\phi \eta $      & $(4.85\pm1.38)\times10^{-5}$ & $(9.07\pm2.52)\times10^{-5}$& $(9.16\pm0.87)\times10^{-5}$ & $(1.98\pm0.16)\times10^{-3}$\\
$\phi \eta'$      & $(2.15\pm0.83)\times10^{-6}$ & $(2.39\pm0.96)\times10^{-6}$& $(1.19\pm0.40)\times10^{-5}$ & $(2.08\pm0.55)\times10^{-5}$\\
$\omega \pi$      & $(3.19\pm0.97)\times10^{-6}$ & $(3.68\pm1.07)\times10^{-6}$& $(3.75\pm2.42)\times10^{-6}$ & $(4.85\pm2.87)\times10^{-6}$\\
$\rho \eta$       & $(2.62\pm0.87)\times10^{-6}$ & $(2.71\pm0.89)\times10^{-6}$& $(2.48\pm1.87)\times10^{-6}$ & $(2.69\pm1.98)\times10^{-6}$\\
$\omega \eta'$    & $(1.41\pm0.26)\times10^{-5}$ & $(2.21\pm0.34)\times10^{-5}$& $(1.13\pm0.74)\times10^{-5}$ & $(1.74\pm1.07)\times10^{-5}$\\
$\rho \eta'$      & $(2.16\pm0.73)\times10^{-6}$ & $(2.19\pm0.74)\times10^{-6}$& $(1.95\pm1.52)\times10^{-6}$ & $(2.04\pm1.56)\times10^{-6}$\\
\midrule[0.5pt]
$\omega K\bar K$  & $(9.92\pm0.04)\times10^{-2}$ & $(2.23\pm0.01)\times10^{-1}$ &                            &                             \\
$\phi K\bar K$    &                              &                              &$(2.02\pm0.01)\times10^{-1}$& $(4.55\pm0.02)\times10^{-1}$\\
\bottomrule[1pt]
\end{tabular}
\end{center}
\end{table*}

For $X(2440)$, our estimations indicate the branching ratio of $X(2440)\to \rho \pi^0$ is up to the order $10^{-2}$, and the one for $X(2440)\to K^{\ast0} \bar{K}^0$ is about one order smaller than the one of $X(2440)\to \rho \pi^0$.
While the branching fractions of the other seven $PV$ channel are even smaller.
As for $X(2680)$, the branching fractions of $K^\ast \bar{K}^0$ channel is of the order $10^{-2}$, which is about one order larger than the one of $X(2680)\to \rho \pi^0$.

Besides the two body $PV$ decay process, the branching ratios of the three body decay processes, $X(2440)\to \omega K\bar{K}$ and $X(2680)\to \phi K\bar{K}$, are also estimated, where $K\bar{K}$ are the daughter particles of $f_0(1710)$.
Our estimations indicate the branching fractions of these three body decay processes are of the order $10^{-1}$.
To data, the decay properties of $X(2440)$ and $X(2680)$ have been poorly measured, we hope our results shown in Table~\ref{tab4} could be checked by future measurements from BESIII, BELLE and LHCb, and probably the future charm-tau factory(FCTF).

\subsection{The branching fractions of $J/\psi \to PV$}
In the mixing scheme, the physical $J/\psi$ state is the mixture of $|c\bar{c}(1S)\rangle$, $|X_{f_0 \omega}\rangle$ and $|X_{f_0 \phi}\rangle$ molecular states as shown in Eq.~(\ref{Eq:Mixing}). Considering the fact that the molecular state $X_{f_0 \omega}/X_{f_0 \phi}$ can decay into a light pseudoscalar and a light vector meson as shown in last subsection, thus one can conclude that the $X_{f_0 \omega}/X_{f_0 \phi}$ molecular components in $J/\psi$ state should also contribute to the processes $J/\psi \to PV$, and the partial widths of $J/\psi \to VP$ satisfy,
\begin{eqnarray}
\label{Eq:Jpsi2PV}
\Gamma(J/\psi \to VP)_{\mathrm{Theory} } &=& \frac{1}{2\;m_{J/\psi}} \int d\Pi_2 \nonumber \\
&& |c_{31} \; \mathcal{M}(X_{f_0\omega} \to VP)\nonumber\\
&& + c_{32}\; \mathcal{M}(X_{f_0\phi} \to VP)\\
&& + c_{33}\; \mathcal{M}(c\bar c(1S) \to VP)|^2 \nonumber.
\end{eqnarray}
Since the mass of $\psi^\prime$ is far above the one of the $X_{f_0 \omega}$ and $X_{f_0 \phi}$ molecular states, $\psi^\prime$ could be considered as a pure $c\bar{c} (2S)$ charmonium state. Then, the partial width of $\psi^\prime \to PV$ should be dominated by the $c\bar{c}$ annihilation. In the present estimation, we estimate the $\Gamma(c \bar c(1S) \to VP)$ by $\Gamma(\psi^\prime \to VP)$ and the ``$12\%$ rule". The terms of $\Gamma(X_{f_0 \omega}\to VP)$ and $\Gamma(X_{f_0 \phi}\to VP)$ are estimated in the same way as the molecular decay. However, in the physical $J/\psi$ state, $X_{f_0 \omega}$ and $X_{f_0 \phi}$ are off-shell, thus, the coupling constants $g_X$ estimated in Eq.~(\ref{Eq:gX}) is not valid and we take $g_X$ as an undetermined parameter. As for the factors $c_{31}$, $c_{32}$ and $c_{33}$, they are determined by the unitary transformation matrix $U$ in Eq.~(\ref{18}), which is,
\begin{eqnarray}\nonumber
\begin{aligned}
c_{31}^2&=0.0107\pm0.0070&,\\
c_{32}^2&=0.0146\pm0.0114&,\\
c_{33}^2&=0.9747\pm0.0050&.\\
\end{aligned}
\end{eqnarray}

For better understand the contributions of different terms in Eq.~(\ref{Eq:Jpsi2PV}), we have the following relations and definitions,
\begin{widetext}
\begin{eqnarray}
\mathcal{B}(c\bar c(1S)\to PV)_{\mathrm{Theory}}&=& {\mathcal{B}(\psi^\prime\to PV)_{\mathrm{Exp}}}/{(13.3\%} \times c_{33}^2), \nonumber \\
\mathcal{B}(J/\psi\to PV)_{\mathrm{Theory}}&=& \mathcal{B}(c\bar c(1S)\to PV)_{\mathrm{Theory}}
+ \mathcal{B}(X\to PV)_{\mathrm{Theory}} +\mathcal{B}(INT\to PV)_{\mathrm{Theory}}, \label{Eq:definitions}
\end{eqnarray}
\end{widetext}
where the $INT$ stands for the interference terms form molecular states with $c\bar c(1S)$.
The estimated branching fractions defined above are listed in Table~\ref{tab5}.
In this table, the branching fraction of the process $\psi^\prime \to \rho^0 \pi^0$ is estimated by $\mathcal{B}(\psi^\prime\to \rho^0\pi^0)=\mathcal{B}(\psi^\prime \to \rho\pi)/3$. The decay processes $\psi^\prime \to \phi \pi$ and $J/\psi \to \phi \pi$ are not included since the branching fraction of these two channels are extremely small.
As for the decay channels $J/\psi(\psi^\prime)\to \rho\eta$, $J/\psi(\psi^\prime)\to \omega\eta^\prime$ and $\psi^\prime \to \rho \eta^\prime$, since from Table.~\ref{tab1} one can find that the three channels satisfy the $12\%$ rule within the error range of experiment. Thus we estimate that the components of $X\to \rho\eta$, $X\to \omega\eta'$ and $X\to \rho\eta'$ contribution in $J/\psi$ is no lager than $10^{-5}$, and our estimations in Table~\ref{tab5} are consistent with the expectations.
\begin{table*}[htb]
\begin{center}
\caption{\label{tab5}
The experimental values of branching ratio of $J/\psi\to PV$ and $X\to PV$ with $X$ stands for both $X_{f_0\omega}$ and $X_{f_0\phi}$.
The uncertainties of $\mathcal{B}(X\to PV)_{\mathrm{Theory}}$ and $\mathcal{B}(INT\to PV)_{\mathrm{Theory}}$ are resulted from $c_{ij}$.
The uncertainties of $\mathcal{B}(c\bar c(1S)\to PV)_{\mathrm{Theory}}$ is from experimental data through ``$12\%$ rule".
}
\renewcommand\arraystretch{1.5}
\begin{tabular}{p{1cm}<\centering p{3cm}<\centering p{3cm}<\centering p{3cm}<\centering p{3.5cm}<\centering p{3.7cm}<\centering }
\toprule[1pt]
Channel~      &~$\mathcal{B}(\psi'\to PV)_{\mathrm{Exp}}$ &~$\mathcal{B}(J/\psi\to PV)_{\mathrm{Exp}}$ ~&~
$\mathcal{B}(c\bar c(1S)\to PV)_{\mathrm{Theory}}$ ~&~ $\mathcal{B}(X\to PV)_{\mathrm{Theory}}/(g_{f_0}g_X)^2$ &
$\mathcal{B}(INT\to PV)_{\mathrm{Theory}}/(g_{f_0}g_X)$\\ \midrule[1pt]
$\rho^0 \pi^0$    & $(1.1\pm0.4)\times10^{-5}$   & $(5.6\pm0.7)\times10^{-3}$   & $(7.82\pm2.93)\times10^{-5}$  & $(3.03\pm2.02)\times10^{-3}$ & $(9.89\pm3.76)\times10^{-4}$ \\
$K^{*+}\bar{K}^-$ & $(1.5\pm0.2)\times10^{-5}$   & $(3.0\pm0.5)\times10^{-3}$   & $(1.06\pm0.15)\times10^{-4}$  & $(3.29\pm2.42)\times10^{-3}$ & $(1.21\pm0.53)\times10^{-3}$\\
$K^{*0}\bar{K}^0$ & $(5.5\pm1.0)\times10^{-5}$   & $(2.1\pm0.2)\times10^{-3}$   & $(3.99\pm0.73)\times10^{-4}$  & $(3.29\pm2.42)\times10^{-3}$ & $(2.34\pm1.02)\times10^{-3}$\\
$\omega\eta$      & $<1.1\times10^{-5}$          & $(1.74\pm0.20)\times10^{-3}$ & $<8.06\times10^{-5}$          & $(1.96\pm1.33)\times10^{-4}$ & $(2.52\pm0.98)\times10^{-4}$  \\
$\phi\eta $       & $(3.10\pm0.31)\times10^{-5}$ & $(7.4\pm0.8)\times10^{-4}$   & $(2.27\pm0.23)\times10^{-4}$  & $(2.10\pm1.57)\times10^{-4}$ & $(4.40\pm1.98)\times10^{-4}$\\
$\phi\eta'$       & $(1.54\pm0.20)\times10^{-5}$ & $(4.6\pm0.5)\times10^{-4}$   & $(1.13\pm0.15)\times10^{-4}$  & $(2.26\pm1.74)\times10^{-6}$ & $(3.29\pm1.55)\times10^{-5}$\\
$\omega\pi$       & $(2.1\pm0.6)\times10^{-5}$   & $(4.5\pm0.5)\times10^{-4}$   & $(1.54\pm0.44)\times10^{-4}$  & $(8.18\pm5.88)\times10^{-8}$ & $(7.30\pm3.12)\times10^{-6}$ \\
$\rho\eta$        & $(2.2\pm0.6)\times10^{-5}$   & $(1.93\pm0.23)\times10^{-4}$ & $(1.61\pm0.44)\times10^{-4}$  & $(6.54\pm4.71)\times10^{-9}$ & $(2.00\pm0.85)\times10^{-6}$\\
%$\phi\pi$        & $<3\times10^{-6}$                & $<9.6\times10^{-8}$                 & $<2.6\times10^{-10}$ \\  \hline
$\omega\eta'$     & $(3.1\pm2.5)\times10^{-5}$   & $(1.89\pm0.18)\times10^{-4}$ & $(2.35\pm1.83)\times10^{-4}$  & $(9.39\pm6.41)\times10^{-6}$ & $(9.60\pm3.13)\times10^{-5}$\\
$\rho\eta'$       & $(1.9\pm1.7)\times10^{-5}$   & $(8.1\pm0.8)\times10^{-5}$   & $(1.39\pm1.25)\times10^{-4}$  & $(2.41\pm1.72)\times10^{-9}$ & $(1.16\pm0.49)\times10^{-6}$\\  \bottomrule[1pt]
%\bottomrule
\end{tabular}
\vspace{0mm}
\end{center}
\end{table*}

In order to better understand the contributions of molecular states to $J/\psi$ hadronic decays, we define the ratio $\mathcal{R}$ as
\begin{eqnarray}
\mathcal{R}&=&\dfrac{\mathcal{B}(J/\psi\to PV)_{\mathrm{Theory}} }{\mathcal{B}(J/\psi\to PV)_{\mathrm{Exp}} } .
\end{eqnarray}
Here we briefly discuss the effective coupling constant $g_X$ and $g_{f_0}$, different from decay processes $X(2680) \to PV$ and $X(2440) \to PV$ in which the effective coupling constant $g_X$ are evaluated through the wave function of the physical state. In $J/\psi \to PV$ process, we set $g_X$ as a free undetermined coupling constants for the  off-shell effect. From the effective Lagrangian in Eq.~(\ref{19}), one has $\mathcal{B}(J/\psi\to f_0(1710)\phi)_{\mathrm{theory} }=g_X^2\;(6.92\pm 5.40)\times10^{-1}$ and $\mathcal{B}(J/\psi\to f_0(1710)\omega)_{\mathrm{theory} }=g_X^2\;(10.38 \pm 6.79)\times10^{-1}$, respectively, with the uncertainties are resulted from $c_{ij}$. Namely, the upper limit of $g_X^2$ is of the order of $10^{-1}$, and thus, one has $g_X^2 g^2_{f_0}\sim 1$.

The ratios $\mathcal{R}$ defined above are listed in Table.~\ref{tab6} with $g_X g_{f_0}$ varies between $0.5\sim1.5$.
\begin{table*}[htb]
\begin{center}
\caption{\label{tab6}
The values of $\mathcal{R}$ with $g_X g_{f_0}$ take $0.5$, $1.0$ and $1.5$ as examples.
}
\renewcommand\arraystretch{1.5}
\begin{tabular}{p{1cm}<\centering p{3cm}<\centering p{3cm}<\centering p{3cm}<\centering  }
\toprule[1pt]
Channel~          &~$g_X g_{f_0}= 0.5$ &~$g_X g_{f_0}= 1.0$ ~&~ $g_X g_{f_0}= 1.5$ \\ \midrule[1pt]
$\rho^0 \pi^0$    & $0.24\pm0.13$      & $0.73\pm0.44$       & $1.50\pm0.94$  \\
$K^{*+}\bar{K}^-$ & $0.52\pm0.31$      & $1.55\pm1.03$       & $3.11\pm2.15$  \\
$K^{*0}\bar{K}^0$ & $1.14\pm0.58$      & $2.89\pm1.71$       & $5.39\pm3.40$  \\
$\omega\eta$      & $0.15\pm0.05$      & $0.30\pm0.14$       & $0.52\pm0.26$  \\
$\phi\eta $       & $0.68\pm0.23$      & $1.19\pm0.53$       & $1.84\pm0.93$  \\
$\phi\eta'$       & $0.28\pm0.06$      & $0.32\pm0.08$       & $0.36\pm0.10$  \\
$\omega\pi$       & $0.35\pm0.11$      & $0.36\pm0.11$       & $0.37\pm0.12$  \\
$\rho\eta$        & $0.84\pm0.25$      & $0.85\pm0.25$       & $0.85\pm0.26$  \\
%$\phi\pi$        & $<3\times10^{-6}$                & $<9.6\times10^{-8}$                 & $<2.6\times10^{-10}$ \\  \hline
$\omega\eta'$     & $1.51\pm1.09$      & $1.80\pm1.21$       & $2.11\pm1.31$  \\
$\rho\eta'$       & $1.73\pm1.55$      & $1.73\pm1.55$       & $1.74\pm1.56$  \\  \bottomrule[1pt]
%\bottomrule
\end{tabular}
\vspace{0mm}
\end{center}
\end{table*}
From Table.~\ref{tab6} we find that our estimated branching fractions of $J/\psi\to \rho^0 \pi^0 $, $J/\psi\to K^*\bar{K}$, $J/\psi\to\phi\eta$, $J/\psi\to \rho\eta$, $J/\psi\to \omega\eta'$ and $J/\psi\to \rho\eta'$ are consistent with the experimental expectations.
The channels $J/\psi\to\omega\eta$ can only meet experimental expectations within $3\sigma$ error ranges. While for $J/\psi\to\phi\eta'$ and $J/\psi\to\omega\pi$, the theoretical value is several times smaller than experimental expectation.

We also note that the experimental data $J/\psi\to f_0(1710)\phi\to \phi K\bar K$ and $J/\psi\to f_0(1710)\omega\to \omega K\bar K$ may have a confinement to $c_{ij}g_X g_{f_0}$ through $f_0(1710)\to K\bar K$ process.
The corresponding coupling constants can be fitted through following relations and equations from Eq.\ref{19},
\begin{widetext}
\begin{eqnarray}
\mathcal{M}(J/\psi\to f_0(1710)\phi\to \phi K\bar K)&=&2g_{f_0} m_K\dfrac{1}{q^2-m_{f_0}^2}\epsilon^{*\mu}_\phi g_{\mu\nu}\epsilon^\nu_X g_X \Lambda_0 \sqrt3 \cos\varphi,\\
\Gamma(J/\psi\to f_0(1710)\phi\to \phi K\bar K)&=&\frac{1}{2\;m_{J/\psi}}\int d\Pi_3 |c_{32}\mathcal{M}(J/\psi\to f_0(1710)\phi\to \phi K\bar K)|^2.
\label{eq2}
\end{eqnarray}
\end{widetext}
By fitting $\mathcal{B}(J/\psi\to f_0(1710)\phi\to \phi K\bar K)_{Exp}=(3.6\pm0.6)\times10^{-4}$~\cite{DM2:1988qci} we can obtain $g_X g_{f_0}=0.09\pm0.02$. And by fitting $\mathcal{B}(J/\psi\to f_0(1710)\omega\to \omega K\bar K)_{Exp}=(4.8\pm1.1)\times10^{-4}$~\cite{DM2:1988qci} we have $g_X g_{f_0}=0.08\pm0.02$. We also find that, when taking the two values into relation $\mathcal{B}(J/\psi\to PV)_{\mathrm{Theory}}$, we can obtain numerical results which are independent of $c_{ij}$, since $c_{ij}$ and $g_X g_{f_0}$ exist in both Eq.\ref{Eq:Jpsi2PV} and Eq.\ref{eq2}.

Now let us discussion the values of $g_X g_{f_0}$ fitted from experiment which are much smaller. Firstly, although $g_X g_{f_0}$ is fitted through $J/\psi$ three hadrons final state experimental data, $g_X$ and $g_{f_0}$ are off shell values, the running of those effective coupling constants are hard to handle. It may effect the numerical values. This point have also been mentioned above, especially for $g_X$.
Secondly, for the value of $g_{f_0}$, the experimental data is lacked, more $f_0(1710)\to PP(VV)$  and $f_0(1710)$ hadronic decays data are needed. Moreover, we hope much more $J/\psi$ hadronic and radiative decays containing $f_0(1710)$ resonance peak will be observed and measured in experiment to fit the effective coupling constants.
Finally, for the $\omega-\phi$ mixing we adopt $\sin\varphi=-0.76$, but its value is not well studied, and this will effective the numerical values.

\section{Summary}

Motivated by the recently observations of a series of light exotic candidates and the anomalous large branching fraction of  $J/\psi\to f_0(1710)\phi/f_0(1710)\omega$,
we suppose that the experimental observed resonances $X(2440)$ and $X(2680)$ are $1^{--}$ state and mainly composed of $X_{f_0\omega}$ and $X_{f_0\phi}$ molecular states, respectively.
Meanwhile, the two molecular states $X_{f_0\omega}$ and $X_{f_0\phi}$ can mix with $c\bar c(1S)$ to form experimental observed states $X(2440)$, $X(2680)$ and $J/\psi$.
In the present work, we first evaluate the mass spectra of $c\bar c$ bound states with the Godfrey-Isgur model~\cite{Godfrey:1985xj}, then the masses of two molecular states $X_{f_0\omega}$ and $X_{f_0\phi}$ are also evaluated within the framework of the OBE model~\cite{Chen:2019uvv,Zhao:2013ffn}. Then we investigate the mixing of $c\bar c(1S)$ bound state with the two molecular states $X_{f_0\omega}$ and $X_{f_0\phi}$ and obtain the mixing parameters through reproducing the experimental data.

Moreover, we estimated the branching fraction of  $X(2680)/X(2440)\to PV$ and the three body final states $K\bar K\phi(\omega)$. Our estimations indicate the results strongly depend on the effective coupling constants $g_{f_0}$ and $g_X$, and the $\omega-\phi$ mixing angle $sin\varphi$. With a certain parameter ranges, the rate of some $J/\psi\to PV$ channels can be naturally understood with the present ansatz.

The experimental data $\mathcal{B}(J/\psi \to f_0(1710)\phi\to \phi K\bar K)$~\cite{DM2:1988qci} and $\mathcal{B}(J/\psi\to f_0(1710)\omega\to \omega K\bar K)$~\cite{DM2:1988qci} are observed in the invariant mass spectrum of $J/\psi\to  \phi K\bar K$ and $J/\psi\to \omega K\bar K$, thus we hope the $f_0(1710)$ resonance will be observed in other three body hadronic decays of $J/\psi$, such as $\mathcal{B}(/\psi\to \phi(\omega)\pi\pi)$, $\mathcal{B}(J/\psi\to \phi(\omega)\eta\eta)$, $\mathcal{B}(J/\psi\to \phi(\omega)\rho\rho)$ and $\mathcal{B}(J/\psi\to \phi(\omega)\omega\omega)$.

Unfortunately, the measurement of the hadronic decays of the resonances is still of large uncertainties which hinder us to draw definite values of the coupling constants, thus, it is still too early to making a assured conclusion about the present ansatz. More precise measurements on those hadronic decays in future will provide crucial test to the present ansatz.

\section*{Acknowledgments}

X. D. Guo is very grateful to K. Chen for helpful discussion. This work is supported by National Natural Science Foundation of China under the Grant Number 11805160, 11747040, 11675082 and Qinglan Project of Jiangsu Province in 2021.

\clearpage

\end{document}